\documentclass[draft,jgrga]{agutex}

\usepackage{lineno}
\usepackage{graphicx}
\usepackage{xcolor}
\usepackage{amssymb}
\usepackage{amsmath}
\usepackage{soul}

\setkeys{Gin}{draft=false}

\authorrunninghead{DOSSMANN ET AL.}

\titlerunninghead{MIXING INDUCED BY A NORMAL MODE }

\authoraddr{Corresponding author: Yvan Dossmann, LEMTA, Universit\'e de Lorraine, CNRS, 
France. (yvan.dossmann@univ-lorraine.fr)}

\begin{document}

%----------------------------------------------------------------------------------------
%	TITLE
%----------------------------------------------------------------------------------------

\title{Mixing and formation of layers by internal wave forcing } 

%----------------------------------------------------------------------------------------
%	AUTHORS AND AFFILIATIONS
%----------------------------------------------------------------------------------------

% Use \author{\altaffilmark{}} and \altaffiltext{}

% \altaffilmark will produce footnote; matching \altaffiltext will appear at bottom of page.

\authors{Yvan Dossmann\altaffilmark{1,2}, Florence Pollet\altaffilmark{1}, Philippe Odier\altaffilmark{1} and Thierry Dauxois\altaffilmark{1}}

\altaffiltext{1}{Univ Lyon, ENS de Lyon, Univ Claude Bernard, CNRS, Laboratoire de Physique, F-69342 Lyon, France }
\altaffiltext{2}{LEMTA, UMR 7563, Universit\'e de Lorraine, CNRS, F-54500 Vandoeuvre-les-Nancy, France}

\keypoints{\item Diagnostics for irreversible mixing induced by internal waves in a linear stratification are studied.
           \item The evolution of stratification and turbulent diffusivity are strongly influenced by the normal mode structure.
           \item The formation of staircases by a layering instability is measured in long term experiments.}

%----------------------------------------------------------------------------------------
%	ABSTRACT
%----------------------------------------------------------------------------------------

% Do NOT include any \begin...\end commands within the body of the abstract.

\begin{abstract}

%Breaking internal waves participate into mixing the ocean interior via turbulent processes. Away from any topography, internal waves can break via self interaction mechanisms such as the triadic resonant interaction. The dynamics of a linear background stratification impacted by internal wave induced mixing is still poorly described. 
%Measurements of mixing rates
%turbulent diffusivities
%mixing efficiency
%dynamics of the background stratification: stair case formation.

%Phillips Posmentier model
%Mixing measured when TRI is weaker: mainly normal mode shear induced mixing

The energy pathways from propagating internal waves to the scales of irreversible mixing in the ocean are not fully described. In the ocean interior, the triadic resonant instability is an intrinsic destabilization process that may enhance the energy cascade away from topographies.

The present study focuses on the integrated impact of mixing processes induced by a propagative normal mode-1 over long term experiments in an idealised setup. The internal wave dynamics and the evolution of the density profile are followed using the light attenuation technique. Diagnostics of the turbulent diffusivity $K_{T}$ and background potential energy $BPE$ are provided. Mixing effects result in a partially mixed layer colocated with the region of maximum shear induced by the forcing normal mode. The maximum measured turbulent diffusivity is 250 times larger than the molecular value, showing that diapycnal mixing is largely enhanced by small scale turbulent processes. Intermittency and reversible energy transfers are discussed to bridge the gap between the present diagnostic and the larger values measured in \citet{dossmann_tri_2016}. 

The mixing efficiency $\eta$ is assessed by relating the $BPE$ growth to the linearized $KE$ input. One finds a value of $\Gamma=12-19\%$ larger than the mixing efficiency in the case of breaking interfacial wave. 

After several hours of forcing, the development of staircases in the density profile is observed. This mechanism has been previously observed in experiments with weak homogeneous turbulence and explained by argument. The present experiments suggest that internal wave forcing could also induce the formation of density interfaces in the ocean.

%by which sharp interfaces can form due to vertical variations of the buoyancy flux
%the formation of staircases is observed for the first time in the case of a heterogenous forcing by an internal wave field.

%using high resolution density measurements 

%The staircases are responsible for large variations in the vertical distribution of turbulent diffusivity. 

% These results could help to refine parameterizations of the impact of low order normal modes in ocean mixing.

%The horizontally averaged turbulent diffusivity Kt (z) and the mixing  efficiency  are assessed.  One finds values up to Kt= 10⁻6 m²/s and  %, with slightly larger values in the presence of TRI. The maximum value for Kt is measured at the position(s) of the maximum shear normal mode shear for both normal modes 1 and 2.

\end{abstract}

%----------------------------------------------------------------------------------------
%	ARTICLE CONTENT
%----------------------------------------------------------------------------------------

% The body of the article must start with a \begin{article} command
% \end{article} must follow the references section, before the figures and tables.

\begin{article}

\section{Introduction} \label{article:intro}

Internal waves are involved in diapycnal mixing processes in the stratified ocean. The role played by breaking internal waves in sustaining the meridional overturning circulation (MOC) is still under debate, which emphasizes the need for an accurate description of their contribution to mixing \citep{ferrari2016turning,mcdougall2017abyssal}. 

Simulations of the MOC require parameterizations of mixing processes that occur at spatial scale much smaller than the spatial resolution. A common parameterization for internal waves induced mixing relies on a turbulent diffusivity $K_T$ which is orders of magnitude larger than the molecular diffusivity $\nu$. To assess turbulent diffusivity in the ocean interior from direct observations, assumptions must be made on the fate of energy transported away by the internal wave field. \citet{osborn_estimates_1980}'s parameterization relates the turbulent diffusivity to the turbulent kinetic energy dissipation $\epsilon$ via the equation
\begin{linenomath*}
\begin{equation}
K_T=\frac{\eta}{1-\eta}\frac{\epsilon}{N^2} \label{eq:article:ktosborn}
\end{equation} 
\label{article:eq:kt}
\end{linenomath*}
introducing the mixing efficiency $\eta$, which is the fraction of the IW energy input participating to irreversible mixing of the flow. In \citet{osborn_estimates_1980}'s seminal paper, the flux Richardson number $Ri_f$, which is the ratio of the turbulent buoyancy flux to the turbulent energy production was used instead of $\eta$. As discussed below, recent studies argued that this definition may account for a reversible component to $K_T$. Hence, \citet{osborn_estimates_1980}'s parameterization has been slightly modified in eq.~\ref{article:eq:kt} to unambiguously evaluate the irreversible diapycnal mixing \citep{mashayek2017efficiency}.

The canonical value of $\eta=1/6$ based on homogeneous turbulence experiments is often used in mixing parameterizations, while it has been observed that $\eta$ varies substantially with the mixing processes \citep{mashayek2017efficiency}.

Idealised studies using laboratory experiments and numerical modelling provide useful insights on mixing processes. The limited number of controlled physical parameters (forcing structure and initial stratification) and high resolution density measurements allow for an extensive description of the effects of mixing and to refine existing parameterizations.

The mixing of an initially linear stratification has been widely studied using homogeneous turbulence experiments \citep{thorpe1982layers,ruddick1989formation,park_turbulent_1994,rehmann2004mean}. These experiments provided diagnostics of the mixing efficiency and permitted to investigate in details the layering mechanism \citep{fincham1996energy,pelegri1998mechanism,holford1999turbulent}, first described by \citet{phillips_turbulence_1972} and \citet{posmentier_generation_1977}. 
Experiments on interfacial waves breaking over a topography have been performed by \citet{hult_mixinga_2011,hult_mixingb_2011}. The wave dynamics was followed using Planar Laser Induced Fluorescence, while the evolution of the background stratification was followed by conductivity probe measurements. A bulk mixing efficiency of $3-8\%$ was found, this relatively small value being attributed to the spatial variability of the mixing events. In recent laboratory experiments a ridge was towed through a linear stratification to model the interaction of a quasi-steady abyssal oceanic flow with a submarine ridge \citep{dossmann_leewaves_2016}. Density measurements using the light attenuation technique showed that local mixing in the wake of the ridge largely overcomes the remote mixing induced by emitted lee waves in a large range of parameters. It implies that local mixing processes induced by a quasi-steady flow may have been underestimated with respect to remote mixing. 

Energy transfers toward the scales of mixing can also occur away from any topography. \citet{Joubaud:PoF:12} and \citet{Bourget:JFM:13} studied the generation of secondary internal waves from a primary mode-1 internal wave via the triadic resonant instability process (TRI) using the synthetic Schlieren technique. Through this instability, two secondary waves, verifying temporal and spatial resonant conditions with the primary wave, are generated. These waves have lower frequency and, in general, smaller wavelengths than the primary wave. Therefore they participate to an energy transfer towards smaller scales, where irreversible mixing through diffusion is more likely to happen. 
For this reason, this mechanism is a potential pathway toward the scales of mixing in the ocean \citep{mackinnon2013parametric}. To assess the contribution of TRI to mixing, velocity-density measurements with the simultaneous use of PIV and PLIF were performed \citep{dossmann_tri_2016}. It allowed to assess the eddy diffusivity field $K_{T,<\rho'w'>}$ based on the vertical buoyancy flux. The TRI process yielded values of $K_{T,<\rho'w'>}$ at least an order of magnitude larger than in the case without TRI. Mixing effects reflected on the evolution of the background density profile, as a progressive decrease of the Brunt-V\"ais\"al\"a was measured in the central part of the flow. 
This evolution, as we will show in this article, will modify the propagation of the waves, questioning the long term behavior of the TRI mechanism.

While parameterizations commonly rely on Reynolds-averaged buoyancy flux as a prognostic value for irreversible mixing properties, recent studies argue that reversible processes such as large scale stirring may also contribute to the buoyancy flux \citep{venayagamoorthy2016flux,zhou_diapycnal_2017}. To evaluate the irreversible component of the buoyancy flux that eventually contributes to raise the center of mass of the flow, it is important to provide a diagnostic of mixing processes based on the evolution of the stratification between subsequent mixing events.

In the present study, we perform long term internal wave forcing experiments in a configuration that favors the TRI process. Contrary to mixing induced by convection processes or breaking large amplitude waves in laboratory experiments, the ratio between the energy available in the forcing and the amount of energy locked up in the ambient initial potential energy is rather small in this experiment, so that forcing by a quasi-linear internal wave field must occur over a much larger duration to induce a measurable evolution of the stratification as the energy fluxes at play are much smaller.  Experiments lasting typically $5000$ forcing periods using high resolution density measurements will follow the impact of internal waves mixing on the long term dynamics of the stratification, and provide important diagnostics for mixing events, such as the mixing efficiency and the turbulent diffusivity. In particular, the comparison between the eddy diffusivity based on velocity-buoyancy correlations and the turbulent diffusivity assessed from the vertical mass flux will allow us to discuss some aspects of reversibility in mixing processes.

Section~\ref{article:setup} introduces the setup and the diagnostics for mixing. The flow dynamics and the evolution of the stratification are described in section \ref{article:results}. Mixing diagnostics based on the background potential energy and turbulent diffusivities are presented in sections~\ref{article:bpe} and~\ref{article:kt}. The formation of staircases in the stratification is discussed in section~\ref{article:staircases}. Conclusions are presented in section~\ref{article:conclusion}.

%-need for parameterizations of mixing induced by iw
%-step: understanding better how they contribute to mixing
%-previous works: study of mixing in academic configurations: interfacial waves:Troy and Koseff, gravity current: Hughes, Prastowo. Lee waves: Dossmann 

\section{Set-up and approach} \label{article:setup}

\subsection{Apparatus}

Experiments are conducted in a $80$~cm long, $17$~cm wide tank. It is filled with a linearly stratified fluid to a depth of $H=32$~cm using the double bucket technique \citep{Oster:CR:63}. The stratifying agent is salinity. The initially uniform Brunt-V\"ais\"al\"a frequency is $N_0\equiv\sqrt{(-g/\rho_0)( d\bar{\rho}/dz)} \approx1$~rad\,s$^{-1}$, where $g=9.8$~m\,s$^{-2}$ is the gravity acceleration,  $\rho_0=1000$~kg\,m$^{-3}$ the reference density, $\bar{\rho}(z)$ is the density profile and $z$ is the vertical coordinate oriented upwards.
Internal waves are forced by a custom-made wavemaker placed on the left side of the tank \citep{gostiaux_novel_2007,mercier_new_2010}. The geometry, amplitude and frequency of the forcing are computer-controlled using a LABVIEW program. A normal mode $1$ consisting in a sinusoidal wavelength with node at mid-depth is forced.

The light attenuation technique infers local density from absorption measurements at a high spatial resolution \citep{allgayer_application_1996,hacker_mixing_1996,sutherland_plumes_2012}. This method has previously been used in a similar geometry to describe the dynamics of internal waves generated in the lee of a moving topography as well as the induced mixing \citep{dossmann_leewaves_2016}. 
The tank is backlit with a white luminescent panel and a small volume of blue food dye is added to the brine drum as a passive tracer for density. Images are recorded with a $14$-bits, $2452\times2054$ pixels CCD Pike camera (Allied Vision), placed at $280$ cm from the tank. The camera is mounted with a $35$~mm lens, producing a $40$ cm by $60$ cm field of view. To optimize the measurement sensitivity, a high-pass optical filter with a cut-off wavelength $550$ nm is added in front of the camera lens and the camera exposure is set to maximize the greyscale level range in the dyed fluid.

The light absorption $A(x,z,t)=\log(I_0/I)$ is measured using the intensity field from a reference image ($I_0 (x,z,t)$) and from an image recorded in the course of the experiment $I (x,z,t)$. The initial absorption profile is calibrated against the initial stratification measured with a conductivity probe. This calibration is then applied at each pixel in the tank to infer cross-tank averaged density anomaly fields from light absorption. The reader can refer to~\citet{dossmann_leewaves_2016} for a detailed description of the light attenuation technique. 

\subsection{Experiments}

Experiments are performed over typical forcing durations larger than $1000$ periods and up to $5000$ periods. These durations allow to quantitatively measure the effects of irreversible mixing on the background stratification. Forcing parameters and the background stratification are chosen to favor the generation of Triadic Resonant Interaction (TRI) that potentially contribute to energy transfers toward mixing scales. The maximum displacement amplitude of the wavemaker is $9$ mm and the forcing period $T$ is $7.5$ s, which corresponds to an initial value for nondimensional forcing frequency of $\omega/N=0.9$.
In a given experiment, the flow is forced during regular sequences of $100$ periods ($\simeq 15$ minutes) to $400$ periods ($\simeq 1$ hour) separated by resting periods of $15$ minutes. Three experiments are carried out with the same forcing parameters and initial Brunt-V\"ais\"al\"a frequency to vary the duration of forcing sequences and the total duration of the experiment. The evolution of stratification does not appear to depend on the duration of the forcing sequences in the present experiments. 
At the end of the resting periods, visible turbulent motions have vanished and the flow is assumed to be at rest. The background stratification is measured to quantify the effects of irreversible mixing. 

From the differences between the rest state density profiles before and after each transit of the ridge we provided diagnostics for irreversible mixing.

%\begin{table}
%	\begin{center}
%		\begin{tabular}{|c|c|c|c|}
 %               \hline
 %               Amplitude & Forcing period T & $\omega/N$ & Total forcing duration \\
 %               \hline
 %               $9$ mm & $7.5$ s & $0.9$ & $20$ hours  \\
 %               \hline
 %               \end{tabular}
 %        \end{center}
 %        \caption{Experimental parameters. The experiment is carried out several times to vary the duration of forcing sequences. The latter are separated by resting sequences to allow for measurements of the background stratification.}
 %       \label{table:1}
%\end{table}

\subsection{Irreversible mixing}
\subsubsection{Background potential energy}
The physical framework of~\cite{winters_available_1995} is adopted to quantify the effects of mixing in terms of changes in the background potential energy of the flow. 
%The density field expresses as $\rho(x,z,t)=\overline{\rho}(z)+\rho'(x,z,t)$, where $\overline{\rho}(z)$ and $\rho'(x,z,t)$ are the background density profile and the density anomaly, respectively.
For a density field $\rho(x,z,t)$ in a volume $V$, the potential energy per unit horizontal area is 
\begin{linenomath*}
\begin{equation}
PE(t)=\frac{1}{A_b}\int_{V}g\rho(x,z,t)z\mathrm{d}V, 
\label{eq:article:pe} 
\end{equation} 
\end{linenomath*}
with $A_b$ the bottom area of the tank. The potential energy is partitioned between the background potential energy $BPE(t)$, the minimum potential energy that can be reached through an adiabatic sorting of the density field at each time, and the available potential energy $APE(t)=PE-BPE$. By construction, $BPE(t)$ can only change through diabatic processes such as irreversible mixing while $APE(t)$ is associated with reversible fluid motions.
The density field is similarly expressed as $\rho(x,z,t)=\overline{\rho}(z,t)+\rho'(x,z,t)$, where $\overline{\rho}(z,t)$ is the slowly varying background density profile and $\rho'(x,z,t)$ is the density anomaly associated with fluid motions.
In the present experiments, the background potential energy is measured at the end of each resting period. Once flow motions have ceased, $APE\simeq0$ and the background potential energy per unit horizontal area is simply expressed as
\begin{linenomath*}
\begin{equation}
BPE(t)=PE(t)=\int_{0}^Hg\overline{\rho}(t,z)z\mathrm{d}z. \label{eq:article:bpe}
\end{equation} 
\end{linenomath*}
The difference 
\begin{linenomath*}
\begin{equation}\label{eq:dBPE}
\Delta BPE(t)=BPE(t)-BPE(0)
\end{equation}
\end{linenomath*}
is the irreversible increase of $BPE$ with respect to the initial stratification. This method has been widely used in laboratory experiments to diagnose the effects of irreversible mixing in both stably \citep{hult_mixinga_2011,hult_mixingb_2011,prastowo_mixing_2008,prastowo_topo_2009,fragoso_mixing_2013,hughes2016mixing,micard2016mixing} and convectively unstable stratified flows \citep{dalziel2008mixing,lawrie2011rayleigh,wykes2014efficient} for which values of mixing efficiencies up to $75\%$ were measured.

\subsubsection{Turbulent Diffusivity}
A key parameter in assessing the effects of mixing is the turbulent diffusivity. In this work, the turbulent diffusivity is diagnosed from the irreversible evolution of the background density profile as follows. The equation of mass conservation is used to derive the vertical density flux $F_{\rho}(z,t)$:
\begin{linenomath*}
\begin{equation}
\frac{\partial \bar{\rho}(z,t)}{\partial t}+\frac{\partial F_\rho(z,t)}{\partial z}=0.		
\label{eq:5}
\end{equation}
\end{linenomath*}
The time derivative of the quasi-linearly evolving background density is assessed using $\frac{\partial \bar{\rho}(z,t)}{\partial t}\simeq \frac{\bar{\rho}(z,t)-\bar{\rho}(z,0)}{t}$.
Hence, one can assess $F_\rho$ by a vertical integration of eq. (\ref{eq:5})
\begin{linenomath*}					
\begin{equation}
F_\rho(z,t) = -\frac{1}{\Delta t} \int_{0}^{z} \big(\bar{\rho}(z',t)-\bar{\rho}_{0}(z')\big) \mathrm{d}z', 
\end{equation}
\end{linenomath*}
and deduce the turbulent diffusivity assuming that the vertical mass flux behaves like a diffusive process
\begin{linenomath*}					
\begin{equation}
K_{T,F_\rho}(z,t) = -\frac{F_{\rho}(z,t)}{\frac{\partial \bar{\rho}(z,t)}{\partial z}}. 				
\end{equation}
\end{linenomath*}
The turbulent diffusivity and $BPE$ evolutions are used as diagnostic for internal wave induced mixing in the remainder of the paper. These two quantities are not independent, as the turbulent diffusivity distribution controls the diabatic transfers within the fluid and hence changes in $BPE$. In fact, \citet{barry2001measurements} relates the average turbulent diffusivity to the rate of $BPE$ change. Here, the distribution of $K_{T,F_\rho}$ is used as a diagnostic for local mixing processes and their link to the internal wave field, while $BPE$ describes the bulk effect on the stratification.

\section{Flow dynamics} \label{article:results}

\subsection{Internal wave field} \label{article:fd}

Propagating density anomalies characteristic of a  mode-1 internal wave develop at the beginning of the forcing as seen in Fig.\ref{fig:densshorttimes}a. After a few forcing periods, the emitted and reflected waves interact to form a quasi-steady wave pattern. Later on, secondary waves issued from the TRI process superimpose to the primary wave field (Fig.\ref{fig:densshorttimes}b). Propagating secondary waves are visible over the whole tank length after $\simeq 50$ T. The flow dynamics is similar to the one captured in \citet{dossmann_tri_2016} using PIV/PLIF combined measurements and an earlier version of the internal wave generator as recalled in Fig. \ref{fig:pivlif}. We focus here on the long term evolution of the background stratification by internal wave induced mixing.

\subsection{Background stratification} \label{article:background}

We first examine an experiment with forcing sequences of $100$ periods duration for which the stratification evolution is shown in Fig.\ref{fig:profilesshorttimes}a. Upon forcing, the background stratification undergoes a regular steepening in the central region B ($-22 $cm$<z<-12 $cm), corresponding to a decrease in the Brunt-V\"ais\"al\"a frequency, while the density gradient remains unchanged in the top and bottom part of the flow. The density of fluid particles in regions A and C respectively increases and decreases owing to an upward mass flux $F_\rho(z,t)$. The measured mass of salty fluid is conserved within 0.2$\%$, which validates the accuracy of density measurements using the light attenuation technique. Note also that the vertical position of the inflection point in the density profile is conserved over time. Region B corresponds to the largest values of the horizontal shear induced by the normal mode-1. Despite the presence of secondary waves propagating over the whole tank, shear-induced turbulence forced by the normal mode structure appears to preferentially induce fluid homogeneization in the central region. Complementary experiments performed with a normal mode 2 (not shown) indicate the presence of two steepening regions colocated with the local maxima in horizontal shear of the forcing mode.

\section{BPE evolution and mixing efficiency} \label{article:bpe}

Turbulent mixing induces a partial homogeneization and a raise of the center of mass of the flow. This reflects in a regular increase of the background potential energy of the fluid observed in Fig.\ref{fig:profilesshorttimes}b, at a typical rate of $3$ mJ/m$^2$/s. An estimate of the mixing efficiency can be provided by assessing the kinetic energy horizontal flux carried into the flow by the forcing normal mode 1. It is defined as

\begin{figure}
                                                \centering\includegraphics[width=0.60\textwidth]{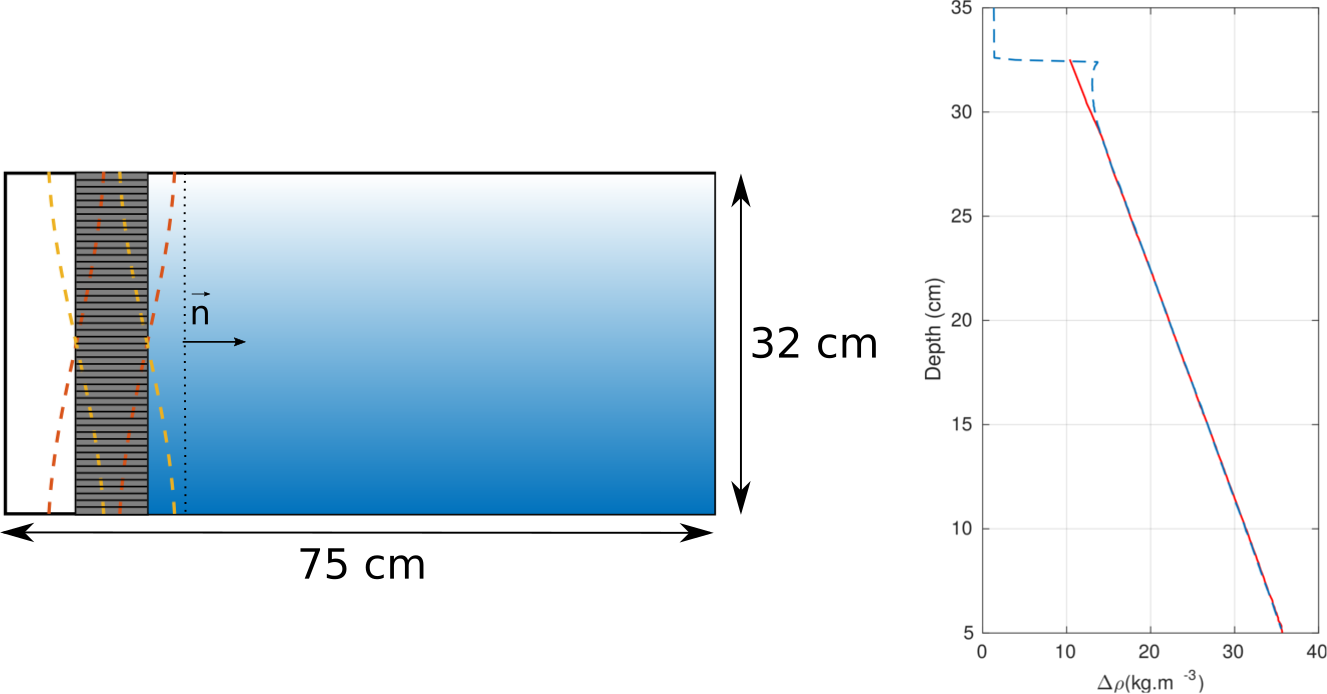}
                                                \caption{Sketch of the experimental apparatus (left). Initial density profile measured with the Light Attenuation Technique (solid red) and conductivity probe (dashed blue)}
                                                \label{fig:sketch}

                                                \end{figure}

\begin{figure}							
								\centering\includegraphics[width=1\textwidth]{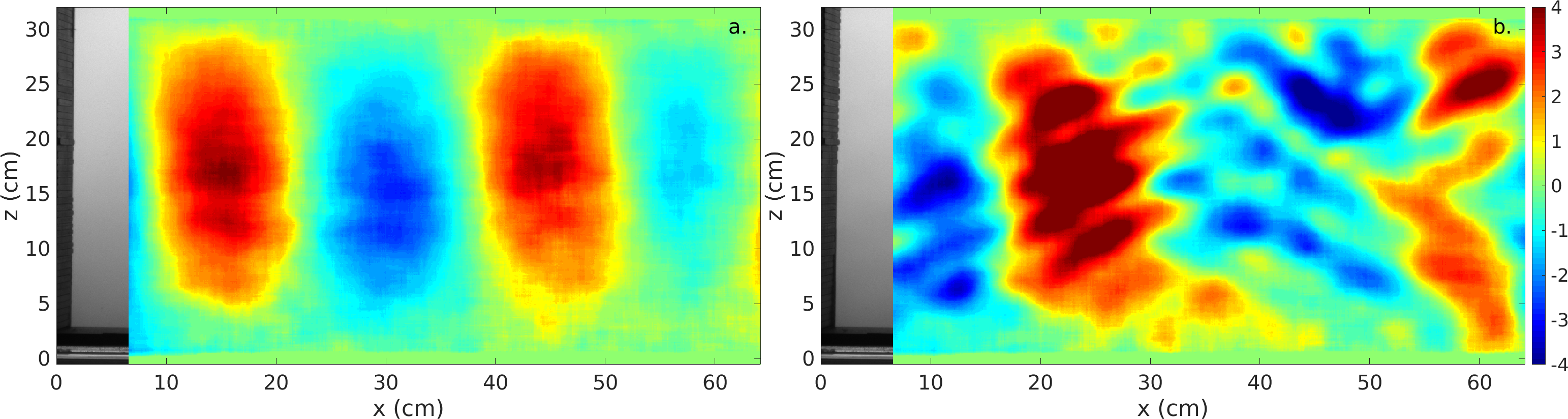}
								\caption{Density anomalies after $27$ T (a.) and $101$ T (b.) measured with the light attenuation technique. The generation of secondary waves via the TRI process is visible in panel b.}
								\label{fig:densshorttimes}
						\end{figure}

\begin{figure}
								\centering\includegraphics[width=0.8\textwidth]{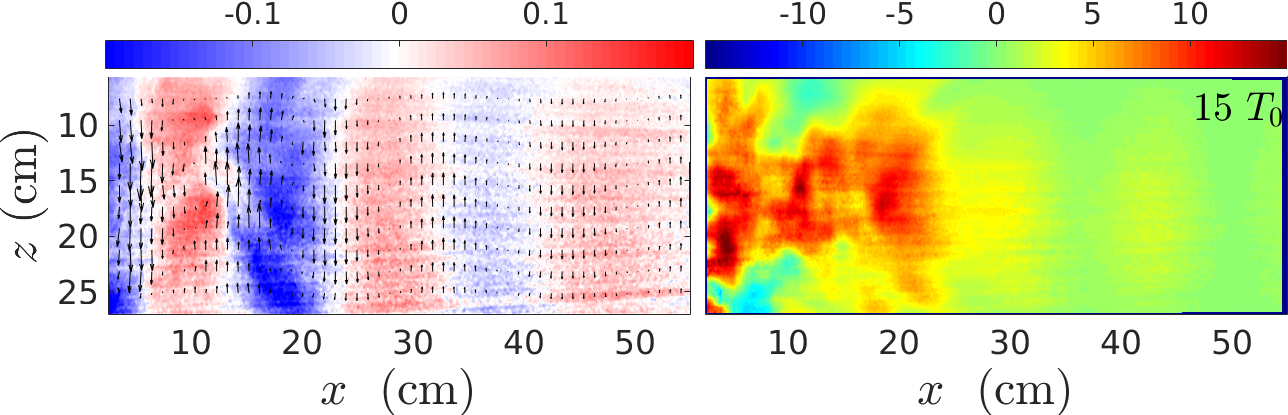}
								\caption{Instantaneous density anomaly (kg/m$^3$) and velocity fields (left). Turbulent diffusivity ($10^{-6}$ m$^2$/s) assessed from the buoyancy/velocity correlation (right). Measurements using PIV/PLIF from \cite{dossmann_tri_2016}}.
								\label{fig:pivlif}
						\end{figure}

\begin{figure}
								\centering\includegraphics[width=0.9\textwidth]{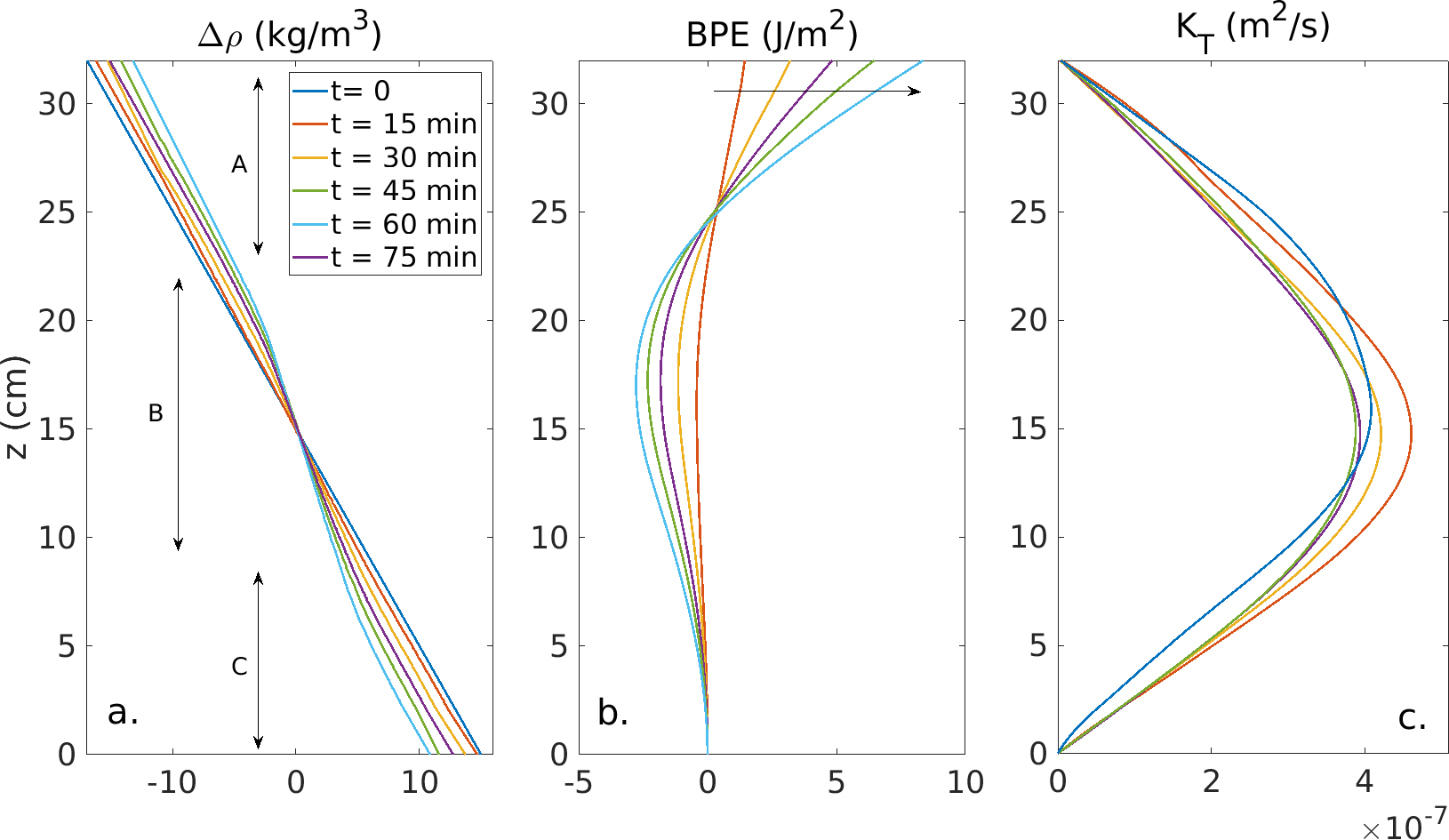}
								\caption{Background density (a), background potential energy (b) and turbulent diffusivity (c) profiles, spaced by sequences of forcing.}
								\label{fig:profilesshorttimes}
						\end{figure}

\begin{linenomath*}
\begin{equation}
F_{KE}=\int_{0}^H<\vec{\pi}>_T\cdot\vec{n} \mathrm{d}z, \label{eq:article:bpe}
\end{equation} 
\end{linenomath*}
where $\vec{\pi}=P'\vec{v}$ is the energy density flux and $<>_T$ denotes a time average over a forcing period T, $P'(x,z,t)$ and $\vec{v}(x,z,t)$ are the pressure anomaly and velocity fields of the flow. The unit vector $\vec{n}$ is along the direction of propagation ($Ox$) of the mode 1 wave, as shown in Fig.~\ref{fig:sketch}. A linear model assuming a normal mode-1 forcing by the generator is used to assess $P'$ and $\vec{v}$. The vertical velocity expresses as
 \begin{equation}
      w(x,z,t) = A_w \sin(mz) \cos(kx-\omega t),
     \label{eq:w}
 \end{equation} 
where $A_w$ is the velocity amplitude, $k$ and $m=\frac{\pi}{H}$ are the horizontal and vertical wavenumbers, respectively. Using the incompressibility and horizontal momentum equations, one obtains

 \begin{align}
     u(x,z,t) & = -A_w\frac{m}{k} \cos(mz) \sin(kx-\omega t), \\
      P'(x,z,t) & = -\rho_0A_w\frac{m\omega}{k^2} \cos(mz) \sin(kx-\omega t).
     \label{eq:u}
 \end{align} 
The kinetic energy flux therefore expresses as
\begin{equation}
F_{KE}=\dfrac{\pi\rho_{0} m \omega }{4k ^{3}}A_w^{2}. \label{eq:article:fke}
\end{equation} 
Using the continuity equation, one can eventually relate $A_w$ to the density anomaly amplitude $A_{\rho'}$
\begin{equation}
A_w = \dfrac{g\omega }{\rho_{0} N^{2}}A_{\rho'},
\end{equation}
and hence
\begin{equation}
F_{KE}=\dfrac{\pi g^2 m \omega^3 }{4 \rho_0N^4k ^{3}}A_{\rho'}^2. \label{eq:article:fkebis}
\end{equation} 

A fraction $\alpha$ of the kinetic energy flux is lost to large scale viscous decay that is not associated with the mixing event, such as viscous damping at the side boundary layers and in the fluid interior~\citep{troy2006viscous}. \citet{hult_mixinga_2011} estimated a value of $\alpha=40-60\%$ in the case of breaking interfacial waves (their Fig.3). This fraction is to be removed from the kinetic energy flux to assess the mixing efficiency.
The dimensionless mixing efficiency expresses as
\begin{equation}
\eta=\dfrac{L_x\times\Delta BPE}{\Delta t}\times\dfrac{1}{F_{KE}(1-\alpha)}, \label{eq:article:gamma}
\end{equation} 
where $L_x$ is the tank length. The quantities $A_{\rho'}$ and $k$ are measured in the density anomaly field at the beginning of the experiment, before the generation of secondary internal waves via TRI.
Typical values of $F_{KE}= 35-50$ mJ/m$^2$/s are measured, which leads to a mixing efficiency $\eta=12-19\%$. This range contains the canonical value used in paramaterizations of internal wave induced mixing and is slightly larger than the value found by \citet{hult_mixinga_2011,hult_mixingb_2011} for mixing induced by an interfacial wave breaking over a topography. 

In an experimental study, \citet{mcewan1983internal} used a similar setup as the present one to assess the mixing efficiency of a standing internal wave in a linearly stratified fluid. The power input was measured using coupled force and displacement transducers connected to the forcing device and the $BPE$ change was measured in a similar manner as in the present work. A mixing efficiency of $\eta=20-32\%$ was assessed. Although these values are of the same order of magnitude in the two sets of experiments, the difference in the forcing devices structure, which consists in a flat paddle in \citet{mcewan1983internal} setup, and the evaluation method for the kinetic enery input can help explaining the discrepancy between the two estimates of mixing efficiency. In particular, the interaction between the reflected waves and the wavemaker are not accounted for in the present experiments, which may lead to a slight overestimation of the kinetic energy input. Friction at the boundaries occur in both setups, but its contribution to the energy losses changes with the (different) tank geometries. This can also lead to a small discrepancy in the evaluation of mixing efficiency.

%A possible explanation for the difference is the evaluation method for the kinetic energy input into the flow. In the present case, the reflected internal wave is  with the present value is the structure of the forcing device which consists in a flat paddle in McEwan's setup

\begin{figure}
								\centering\includegraphics[width=0.7\textwidth]{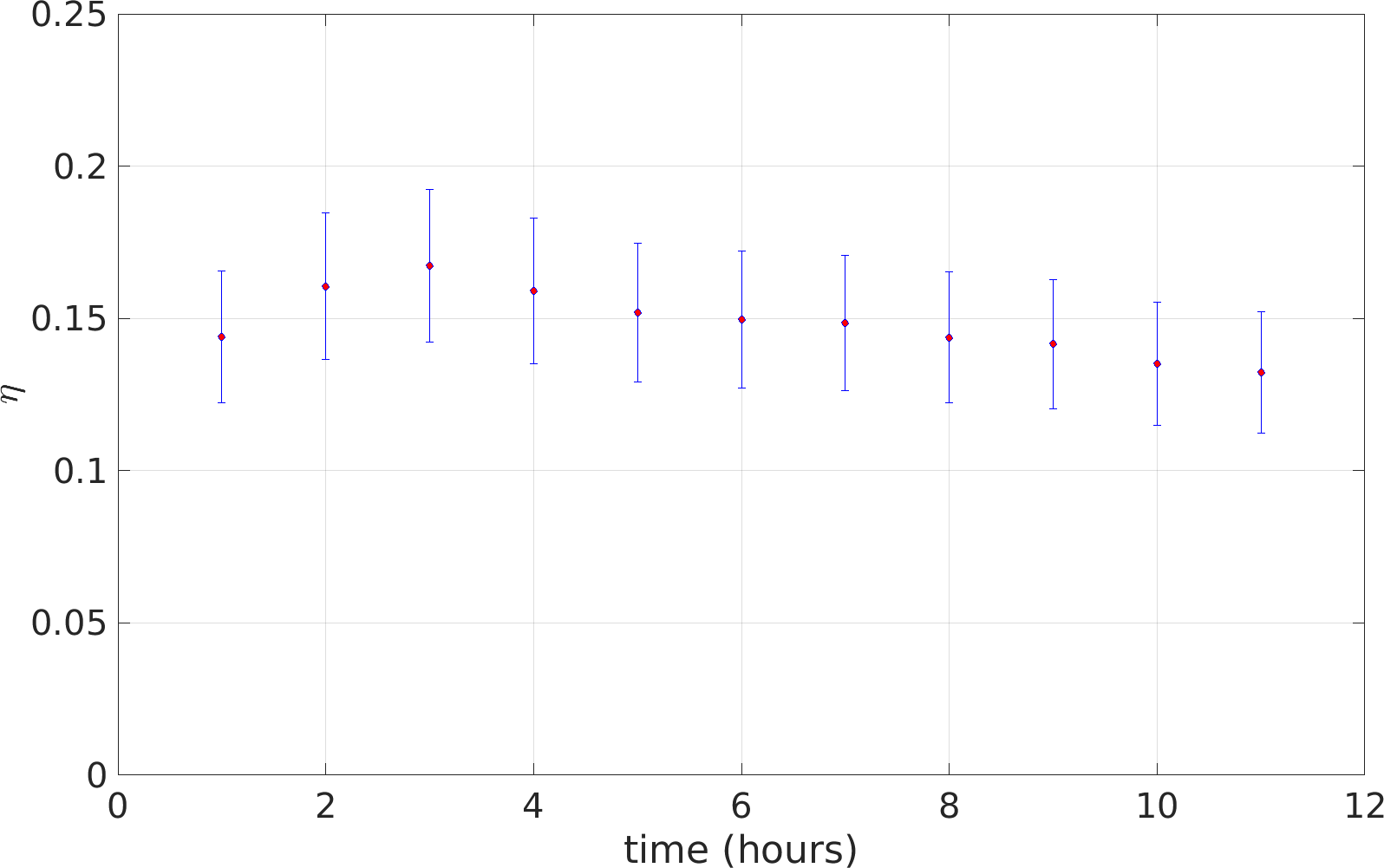}
								\caption{Time series of mixing efficiency.}
								\label{fig:gamma}							                                             
                                                \end{figure}

\begin{figure}								\centering\includegraphics[width=0.4\textwidth]{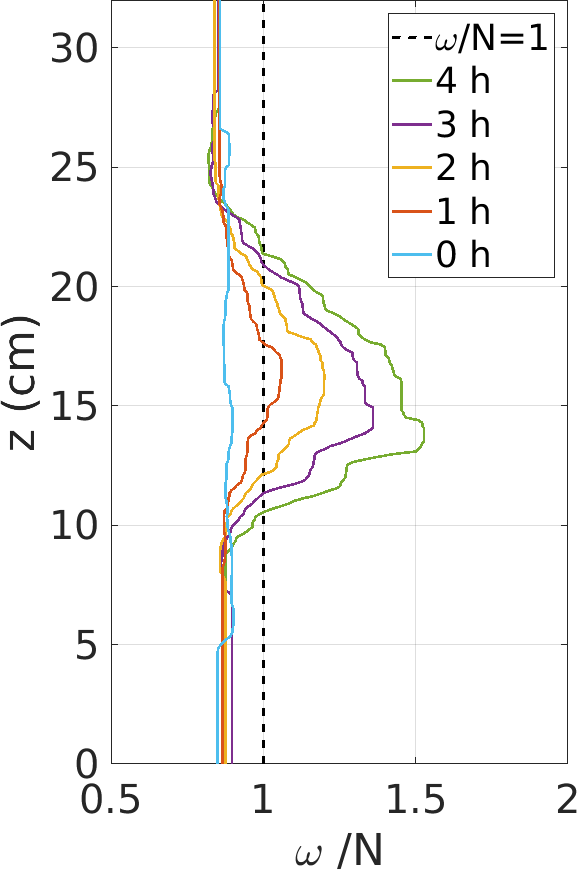}
								\caption{Time series of the $\omega/N$ profile. Shear-induced mixing causes a progressive spreading of the central region where internal waves are evanescent.}
								\label{fig:derivedN}							                                             
                                                \end{figure}

\begin{figure}
								\centering\includegraphics[width=0.5\textwidth]{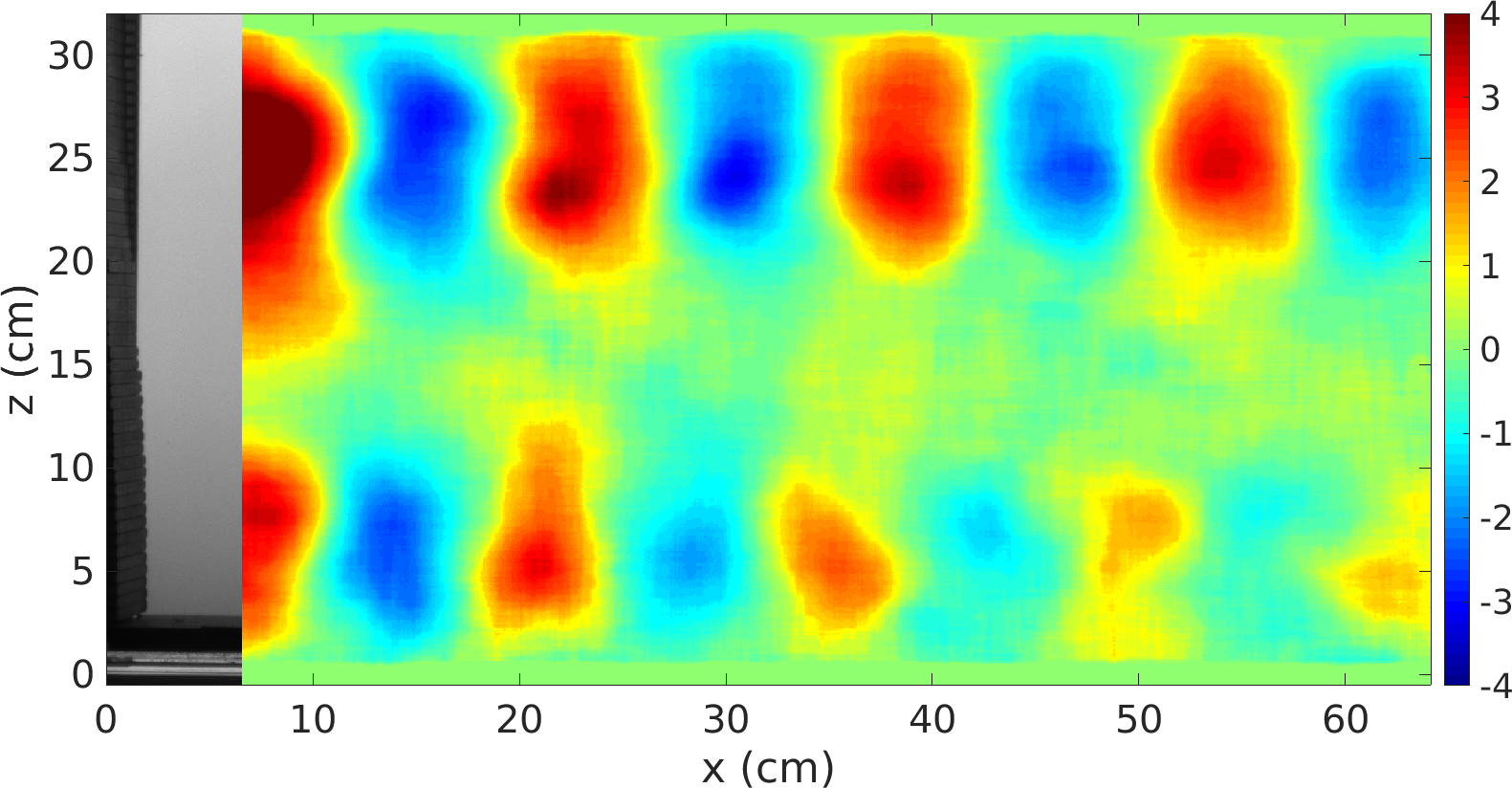}
								\caption{Density anomalies after $2000$ T measured with the light attenuation technique. Internal waves are confined to the top and lower part of the flow.}
								\label{fig:denslongtime}
						\end{figure}

After a moderate increase during the first two hours of forcing, $\eta$ undergoes a regular weak decay as seen in Fig~\ref{fig:gamma}. 
This decay can be explained by the progressive decrease of the Brunt-V\"ais\"al\"a frequency in the center of the tank, at a measured rate of $0.1$~rad/s/hr. After around $350$ T, the forcing frequency $\omega$ overcomes $N$ in a central region that extends upwards and downwards with time, as shown in Fig. \ref{fig:derivedN}. Hence, the large scale mode 1 forcing is limited to the two narrowing regions A and C, while it is evanescent in the increasing central region B, as shown in Fig.\ref{fig:denslongtime}. On the other hand, the smaller frequency secondary waves could still be transporting some of the energy through region B \citep{ghaemsaidi2016nonlinear}. Once the evanescent region is formed, the amount of mixing present in the central region is reduced, but does not vanish completely due to secondary waves. The progressive widening of the evanescent region progressively diminishes the mixing efficiency in the flow with increasing time.

%Hence, propagating internal waves are limited to the two narrowing regions A and C, while they are evanescent in the increasing central region B, as shown in Fig.\ref{fig:denslongtime}. Once the evanescent region is formed, internal waves cannot tranport the largest horizontal shear (located at mid-depth for the normal mode-1) toward the fluid interior

\section{Turbulent diffusivities} \label{article:kt}

The turbulent diffusivity profile is stable over the total duration of the experiment. It reaches $0$ at the top and bottom boundaries due to the no-flux condition, and increases symmetrically from regions A and C to region B, where it reaches a value of $5.10^{-7}$m$^2$/s. The maximum turbulent diffusivity is two orders of magnitude larger than the molecular value for salt diffusion, indicating that turbulent processes are prominent in mixing the stratified fluid.

The turbulent diffusivity profile and maximum value are fairly comparable to the ones measured by~\citet{mcewan1983internal} in a similar configuration (their Fig.4).

The value of $5.10^{-7}$m$^2$/s is however an order of magnitude smaller than the maximum values of $K_{T,<\rho w>}$ measured using buoyancy/velocity correlations with similar forcing parameters in \citet{dossmann_tri_2016}. This discrepancy indicates that the two definitions of $K_T$ are not fully equivalent. Two possible causes can be suggested to explain this difference. 

First, it was shown in \citet{dossmann_tri_2016} that the spatial distribution of the regions of large $K_T$ was not uniform throughout the tank. In addition, this distribution also evolved with time. Therefore, this time intermittency and spatial heterogeneity in $K_{T,<\rho w>}$ can lead, through an integrated effect, to irreversible mixing that is smaller than the instantaneous value.

%$K_{T,<\rho w>}$ as defined potentially transports some reversible contributions such as wave induced stirring. In particular,

Second, the buoyancy flux, as defined, potentially transports a reversible component. In fact the largest values of $K_{T,<\rho w>}$ are measured in the case where the triadic resonant generation is favored, leading to the oblique propagation of secondary waves with spatial scale smaller than the forcing scale. These waves participate in the direct energy cascade toward the scales of mixing and contribute to enhance $K_{T,<\rho w>}$, by comparison with the case of weak TRI \citep{dossmann_tri_2016} However, while a fraction of the secondary waves energy participates to irreversible mixing, some of this energy is involved in reversible transfers between the potential and kinetic energy compartments of the flow \citep[e.g.][]{hughes_available_2009,venayagamoorthy2016flux,zhou_diapycnal_2017}, such as wave induced stirring. These reversible transfers occur at spacetime scales that are different from the forcing scales over which the buoyancy/velocity correlations are averaged and are visibly accounted for in $K_{T,<\rho w>}$. A simple scaling argument permits to assess a scale $l$ for irreversible mixing processes. Assuming a timescale $\omega^{-1}$, one obtains $l\sim\sqrt{K_{T,F_{\rho}}/\omega}\sim1$ mm. This lengthscale is smaller than the secondary wave wavelength by at least one order of magnitude. Hence, the buoyancy fluxes associated with the TRI process occur at scales larger than the ones of irreversible mixing. Consequently, the eddy diffusivity based on these buoyancy fluxes calculated in \citet{dossmann_tri_2016}  contains both a reversible and irreversible contribution, whose partition is still to determine in future works.

The effects of intermittency and reversibility together contribute to the smaller values of $K_{T,F_{\rho}}$ compared to $K_{T,<\rho w>}$. Hence, $K_{T,<\rho w>}$ and $K_{T,F_{\rho}}$ provide complementary information on mixing processes. The former is an instantaneous proxy for the dynamics of stirring at scales larger than irreversible mixing, some of which eventually leading to irreversibly mix the stratified flow. The latter is a diagnostic of the effects of an irreversible mixing event on the background stratification.

The evolution of the background stratification together with the turbulent diffusivity profile highlights that the secondary waves generated by TRI, despite enhancing $K_{T,<\rho w>}$, may not be the primary process leading to irreversible mixing. If so, one could expect a homogeneous value of $K_{T,F_{\rho}}$ (or analogously a homogeneous decrease in $N$) occuring over the tank depth. As discussed before, the colocation of the forcing mode largest horizontal shear stress and the region of intense mixing indicates that the forcing structure controls the irreversible evolution of the stratification.

\section{Staircases formation} \label{article:staircases}

The dynamics of mixing is strongly controlled by the forcing internal wave structure over durations up to $3000$ T, as shown by the strong correlation between the largest values of vertical mass flux and the horizontal shear of the forcing normal mode-1. In order to study the longer term effects of mixing on the stratification, in several experiments the flow was forced over longer durations up to $5000$ T.

%A continuous widening of the mixed central region until the fluid reached a homogeneous density over the whole depth was expected. 

\begin{figure}
								\centering\includegraphics[width=0.4\textwidth]{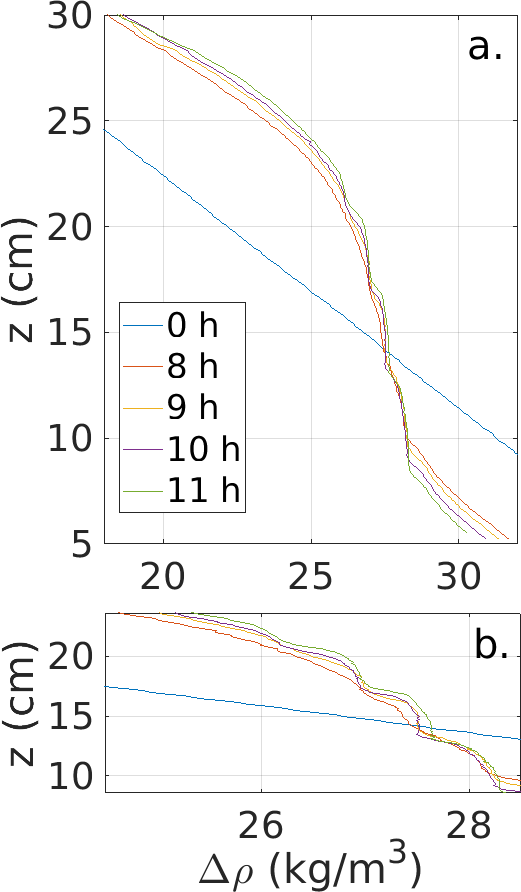}
								\caption{a. Formation of staircases in the density profile after long term forcing. The first set of staircases appears in the central part of the flow, and subsequent staircases form above and below the first set. b. Zoom in the staircase formation region.}
								\label{fig:marche}
						\end{figure}

The result is displayed in Fig.\ref{fig:marche}a and b which show that the central region extends and remains surrounded by two linearly stratified regions at the top and bottom of the flow after 8 hours of forcing. In the mixed region, local gradient increases in the density profile, regularly spaced by $3.5$ cm start forming after around eight hours ($3000$ T). The first pair of perturbations form around mid-depth ($z=13.5$ cm and $z=17$ cm), and subsequent pairs appear above ($z=20.5$ cm) and below ($z=9.5$ cm) the first pair. The local gradients strenghten with time, evolving in a series of interfaces of $5$ mm thickness in the stratification profile, separated by quasi-homogeneous layers of $3$ cm. After $12$ hours of forcing, the typical density jump in the staircases is $0.5$ kg/m$^3$. The stratification stays linear in the regions A and C, with a value of $N$ similar to the initial value. The evaporation process at the free surface and no flux condition at the tank base has generated two mixed layers of centimetric height at the boundaries of the stratified regions (not shown in Fig.8) which do not seem to interact with the stratification in the fluid interior.

The staircases formation strongly reminds of the mechanism independently suggested by \citet{phillips_turbulence_1972} and \citet{posmentier_generation_1977}, for which stable step-like structures can form in an initially linear stratification perturbed by turbulent processes. \citet{linden1979mixing} provided a review of these ideas in the light of laboratory experiments on mixing.

Note that double-diffusion processes are another potential cause for the formation of interfaces in the ocean. The stratifying agent is salinity in the present experiments, while the water temperature remains constant within $\pm 0.1^{\circ}$C, which permits to exclude double diffusion as a cause for layering.

%\yvan{It is based on the combined effects of stratification and the intensity of turbulence, described by the Richardson number Ri, on the vertical mass flux $F_\rho(z,t)$. On the one hand, for small values of $Ri$ diapycnal mixing is inhibited by the small initial density gradient. On the other hand, for large values of $Ri$, the stratification is too strong for vertical motions to develop efficiently and the mass flux is again limited. Consequently there must exist a critical value $Ri_c$ for which the vertical mass flux reaches a maximum. }

%\yvan{When turbulence overcomes the restoring buoyancy force ($Ri<Ri_c$), an initial perturbation to the density profile is smoothened by diapycnal mixing. Conversely, in the region of "weak turbulence" ($Ri>Ri_c$), the decreasing mass flux with $Ri$ can lead to a series of convergence/divergence of the mass flux in the vicinity of a perturbation. The latter can amplify and lead to the development of staircases in the density profile. }

The Phillips-Posmentier mechanism is based on the assumption that the mixing intensity has a non monotonous evolution with the Richardson number, $Ri$, which compares stratification effects to turbulence intensity. This assumption can be justified considering that at small $Ri$, there is little stratification, therefore little to mix, while at large $Ri$, the stratification dominates over turbulence, preventing mixing. For this reason, one expects the existence of an optimal (or critical) value of $Ri$, $Ri_{c}$, for which mixing is maximum, decreasing to zero at both limits of small and large $Ri$. In the region of "weak turbulence" ($Ri>Ri_c$), mixing intensity decreases with increasing $Ri$, so that when a local perturbation to the initially smooth density profile exists, the region of the perturbation where density gradient is lower than ambient (lower $Ri$) will undergo more mixing, resulting in a tendency towards even lower density gradient, while the region of the perturbation where density gradient is larger than ambient (larger $Ri$) will undergo less mixing, therefore keeping a strong gradient. The overall result will be to amplify the perturbation, leading to the development of staircases in the density profile.

The Phillips-Posmentier mechanism has been used to describe layering in several experiments, for which mixing was induced by homogeneous stirring of the flow using traversing vertical rods \citep{ruddick1989formation,park_turbulent_1994} or grids \citep{thorpe1982layers,rehmann2004mean}. 

In the former experiments, homogeneous turbulent mixing induces an important growth of the mixed layer at the top and bottom of the flow, followed by the appearance of staircases spread over the linearly stratified region. The staircases can merge and decay, and the stratification eventually ends up in two mixed layers of constant densities, issued from the continuous growth of the top and bottom mixed layers. Since these experiments were performed, some extended mechanisms have been proposed to explain the formation of layers. The dynamics of their generation is not only controlled by the bulk Richardson number as described in the original mechanism, but also forced by local dynamical structures, such as vortical mixing  \citep{holford1999turbulent,fincham1996energy} or localized regions of intense mixing in shear flows \citep{pelegri1998mechanism}. \citet{thorpe2016layers} offers a review of the various generation mechanisms for layer generation, arguing that the Phillips-Posmentier mechanism can sustain layers forced by other processes. The conditions for formation and stability of a layer have been recently refined relying on numerical studies in a stratified shear flow \citep{zhou_diapycnal_2017,taylor2017multi}. Precise stratification budgets have allowed for the identification of nondimensional parameters for layer formation.
 
In the present case, contrary to previous experiments, the intensity of turbulent mixing varies in the vertical direction, as it depends on the structure of the forcing normal mode (sections \ref{article:bpe} and \ref{article:kt}).
The first set of staircases forms in the central region B of the flow where the largest shear stress is induced, but where turbulence is still sufficiently weak for the staircases instability to develop.
The second set of staircases develops more slowly, as it is generated in a region of weaker shear stress than the first set. 
An important difference compared to the experiments using homogeneous forcing is that the vertical extent of the top and bottom mixed layers remain narrow in the present case. The boundary mixed layers are not affected by turbulent mixing over similar forcing durations as in the homogeneous mixing experiments ($\approx 10$ hours). In fact, internal waves induced shear stress is the smallest at the two boundaries and hence does not enhance the mixed layers growth.

We conclude that the dynamics of staircase formation is strongly sensitive to the structure of the forcing. For the first time in laboratory experiments, the generation of staircases in a linearly stratified fluid is initiated by an internal wave field. The internal wave energy cascades down to the scales of turbulent mixing through local wave overturning and contributes to generate and sustain the layers.

% le nombre de Richardson est environ 100 pour le mode normal linéaire, les gradients ne sont pas suffisants en soi pour le mélange, il faut d'autres processus

\section{Conclusions} \label{article:conclusion}

Mixing induced by a forcing normal mode has been investigated in long term experiments. The dynamics of the internal wave field was followed using the light attenuation technique, in a configuration favoring the generation of secondary waves via the triadic resonant interaction process. The latter are responsible for an enhancement of the eddy diffusivity $K_{T,<\rho w>}$ as measured in \citet{dossmann_tri_2016}.
Secondary waves of smaller spatial scale were indeed observed after $\simeq 50$ T, indicating energy transfers toward smaller spatial scales. The progressive decrease of the background Brunt-V\"ais\"al\"a frequency in the central region B is an evidence of irreversible mixing quantified by $K_{T,F_{\rho}}$. The largest value of $K_{T,F_{\rho}}$ is colocated with the maximum velocity shear of the forcing internal waves, which indicates that the structure of the normal mode controls the vertical distribution of mixing. This idea is further supported by the dynamics of staircase formation discussed in section \ref{article:staircases}. Although the formation of layers appears to be induced by a mechanism analogous to Phillips-Posmentier, some differences in the mixed layer growth rate and the order of layer formation with respect to the case of homogeneous mixing are highlighted \citep{ruddick1989formation,park_turbulent_1994}. They can be explained by the vertical heterogeneity of the normal mode shear stress in the present experiments. A precise evaluation of the different terms in the stratification budget with a similar approach as in \citet{zhou_diapycnal_2017} is needed to quantitatively describe the relation between internal wave induced mixing and the generation of layers. Of particular interest is the link between the formation of a forbidden zone for internal wave propagation and the emergence of layers.

The discrepancy between $K_{T,F_{\rho}}$ and $K_{T,<\rho w>}$ underscores the important questions of the irreversibility of turbulent buoyancy fluxes in mixing processes and their accurate assessments. Both quantities are useful in the study of mixing by waves, since one brings a local assessment of the  stirring dynamics, while the other diagnoses the effects of mixing on the global energy balance. Future experimental studies combining these two measurements could help to better predict the effect of turbulent buoyancy fluxes on the background stratification.

%						\begin{figure}
%						\begin{center}
 %                                               \begin{minipage}{0.94\textwidth}
  %                                              \begin{tabular}{cc}
   %                                             \includegraphics[width=0.60\textwidth]{figures/sketch5.png} &
    %                                            \includegraphics[width=0.21\textwidth]{figures/probelat.pdf} \\
     %                                           \end{tabular}
      %                                          \caption{Sketch of the experimental apparatus (left). Initial density profile measured with the Light Attenuation Technique (solid red) and conductivity probe (dashed blue)}
       %                                         \label{fig:sketch}
        %                                        \end{minipage}
         %                                       \end{center}
          %                                      \end{figure}
                                                
\begin{acknowledgments}
We thank Ross Griffiths, Sylvain Joubaud, Antoine Venaille and Kraig Winters for insightful discussions, and Pascal Metz for the technical backup with the wavemaker. We thank two anonymous reviewers for their relevant suggestions. This work was supported by the LABEX iMUST (ANR-10-LABX-0064) of Universit\'e de Lyon, within the program ``Investissements d'Avenir'' (ANR-11-IDEX-0007), operated by the French National Research Agency (ANR). Data are archived at the PSMN server at the Laboratoire de Physique (ENS Lyon) and available through registration.
\end{acknowledgments}

%----------------------------------------------------------------------------------------
%	BIBLIOGRAPHY
%----------------------------------------------------------------------------------------

\bibliographystyle{agufull08}
\bibliography{biblio_complete2}

\end{article}

%----------------------------------------------------------------------------------------
%	FIGURES AND TABLES
%----------------------------------------------------------------------------------------

%% Enter Figures and Tables here:
%
% DO NOT USE \psfrag or \subfigure commands.
%
% Figure captions go below the figure.
% Table titles go above tables; all other caption information should be placed in footnotes below the table.
%
%----------------
% EXAMPLE FIGURE
%
% \begin{figure}
% \noindent\includegraphics[width=20pc]{samplefigure.eps}
% \caption{Caption text here}
% \label{figure_label}
% \end{figure}
%
% ---------------
% EXAMPLE TABLE
%
%\begin{table}
%\caption{Time of the Transition Between Phase 1 and Phase 2\tablenotemark{a}}
%\centering
%\begin{tabular}{l c}
%\hline
% Run  & Time (min)  \\
%\hline
%  $l1$  & 260   \\
%  $l2$  & 300   \\
%  $l3$  & 340   \\
%  $h1$  & 270   \\
%  $h2$  & 250   \\
%  $h3$  & 380   \\
%  $r1$  & 370   \\
%  $r2$  & 390   \\
%\hline
%\end{tabular}
%\tablenotetext{a}{Footnote text here.}
%\end{table}

% See below for how to make sideways figures or tables.

%\begin{figure}
%\begin{center}
%\includegraphics[width=\linewidth]{./figures/schema_exp_3.pdf}
%\caption{Experimental setup (not to scale). The tank is filled with a linear salinity gradient (inset at right). Red dye is a passive tracer of salt. The rectangle in green broken line (centre) indicates the area of density field measurements and the shaded green rectangle (right) indicates the location of shadowgraph photographs.}
%\label{fig:article:stereogp}
%\end{center}
%\end{figure}

\end{document}